\documentclass[aps,prl,10pt,twocolumn,superscriptaddress,amsmath,amssymb,longbibliography,floatfix]{revtex4-2}
\usepackage[english]{babel}
\usepackage{graphicx,bbm}
\usepackage[utf8]{inputenc}
\usepackage{float}
\usepackage{multirow}
\usepackage[colorlinks,linkcolor=blue,citecolor=blue,urlcolor=blue]{hyperref}

\begin{document} 
\title{Photoinduced switching of orbital-selective Mott states}
\author{M. Lisiecki}
\affiliation{Institute of Theoretical Physics, Faculty of Fundamental Problems of Technology, Wroc{\l}aw University of Science and Technology, 50-370 Wroc{\l}aw, Poland}
\author{P. Werner}
\affiliation{Department of Physics, University of Fribourg, 1700 Fribourg, Switzerland}
\author{J. Herbrych}
\affiliation{Institute of Theoretical Physics, Faculty of Fundamental Problems of Technology, Wroc{\l}aw University of Science and Technology, 50-370 Wroc{\l}aw, Poland}

\date{\today}

\begin{abstract}
We investigate light-driven inversion of orbital selectivity in a two-orbital Hubbard--Kanamori model initially in an orbital-selective Mott phase. Using the matrix-product-state we show that tailored pulse excitations drive interorbital charge transfer that nearly reverses the orbital occupations. Single-particle spectral snapshots exhibit changes consistent with an interchange of the localized and itinerant characters of the orbitals. We analyze the relevant one- and two-photon excitation channels  and show that chirping around the lower-frequency resonance enhances the transfer. These results establish a microscopic route to reversing orbital selectivity and accessing long-lived nonequilibrium states in multiorbital systems.
\end{abstract}
\maketitle

{\it Introduction \textemdash}
Strong electronic correlations in multiorbital transition-metal compounds give rise to a wealth of quantum phases beyond the reach of single-band models~\cite{Imada1998,Georges2013}. The Hund coupling plays a central role here: it governs the degree of correlation in metallic states far from any Mott insulator and can strongly differentiate the behavior of individual orbitals~\cite{Georges2013,deMedici2011}. When orbitals differ sufficiently in bandwidth, filling, or crystal-field energy, this differentiation becomes extreme: some orbitals localize into a Mott-insulating state while others remain itinerant, realizing an orbital-selective Mott phase (OSMP)~\cite{Anisimov2002,Koga2004,Liebsch2007,Werner2007,deMedici2009}. Originally proposed for the ruthenate Ca$_{2-x}$Sr$_x$RuO$_4$~\cite{Anisimov2002}, orbital selectivity is now regarded as a key feature of iron-based superconductors and other Hund-metal compounds~\cite{Lechermann2005,Yi2013,deMedici2014,Yu2013,Georges2013}. The concept also extends to reduced dimensions, where for example quasi-one-dimensional iron-based ladders offer a setting to investigate orbital-selective correlations~\cite{Caron2012,Takahashi2015,Yamauchi2015,Mourigal2015}.

In parallel, ultrafast light pulses have emerged as a versatile tool for manipulating correlated phases out of equilibrium~\cite{Basov2017,delaTorre2021,Aoki2014}. Intense fields can melt Mott insulators and photodope carriers across the charge gap~\cite{Aoki2014,Eckstein2011}, drive transient or long-lived metastable states inaccessible in equilibrium~\cite{Stojchevska2014,Li2018,Kaneko2019,Ray2024}, and engineer effective Hamiltonians through periodic driving~\cite{Oka2019}. Crucially, these protocols provide control knobs - pulse frequency, amplitude, and shape - that can act selectively on specific electronic degrees of freedom.

Most studies of photoexcited Mott systems have focused on the destruction of the insulating state, through photodoping~\cite{Aoki2014,Murakami2025} and the transient collapse or filling of the Mott gap~\cite{Roach1971,Morrison2014,Chen2024,Okamoto2010,Eckstein2011}. Multiorbital systems offer an additional avenue for control: photoexcitation can redistribute charge among orbitals and produce local excitations governed by the Hund coupling~\cite{Gillmeister2020,Ray2023,Ray2024,Lojewski2024,Rajpurohit2026}. Such interorbital charge transfer is particularly consequential near an orbital-selective Mott transition, where changes in orbital filling can destabilize Mott localization in an individual orbital~\cite{Jakobi2013}. This raises the possibility of controlling not only the transport properties and strength of correlations, but also the orbital character of the itinerant and localized electrons. However, whether light can switch the localized and itinerant orbitals of an OSMP remains an open question.

In this Letter, we demonstrate that a tailored light pulse can dynamically invert an OSMP. Using matrix-product-state (MPS) simulations of a two-orbital Hubbard-Kanamori model~\cite{Herbrych2019,Herbrych2020-1,Herbrych2020-2,Herbrych2021}, we show that a pump pulse transfers charge from the initially itinerant orbital to the Mott-localized one, nearly reversing the orbital occupations and interchanging their characters. We identify the one- and two-photon excitation channels responsible for the transfer and show that the resulting inverted OSMP is long-lived, persisting well beyond the pulse. Our results establish tailored light fields as a route to controlling orbital-selective correlations and creating long-lived nonequilibrium states in multiorbital systems.

\begin{figure*}[htb!]
\includegraphics[width=\textwidth]{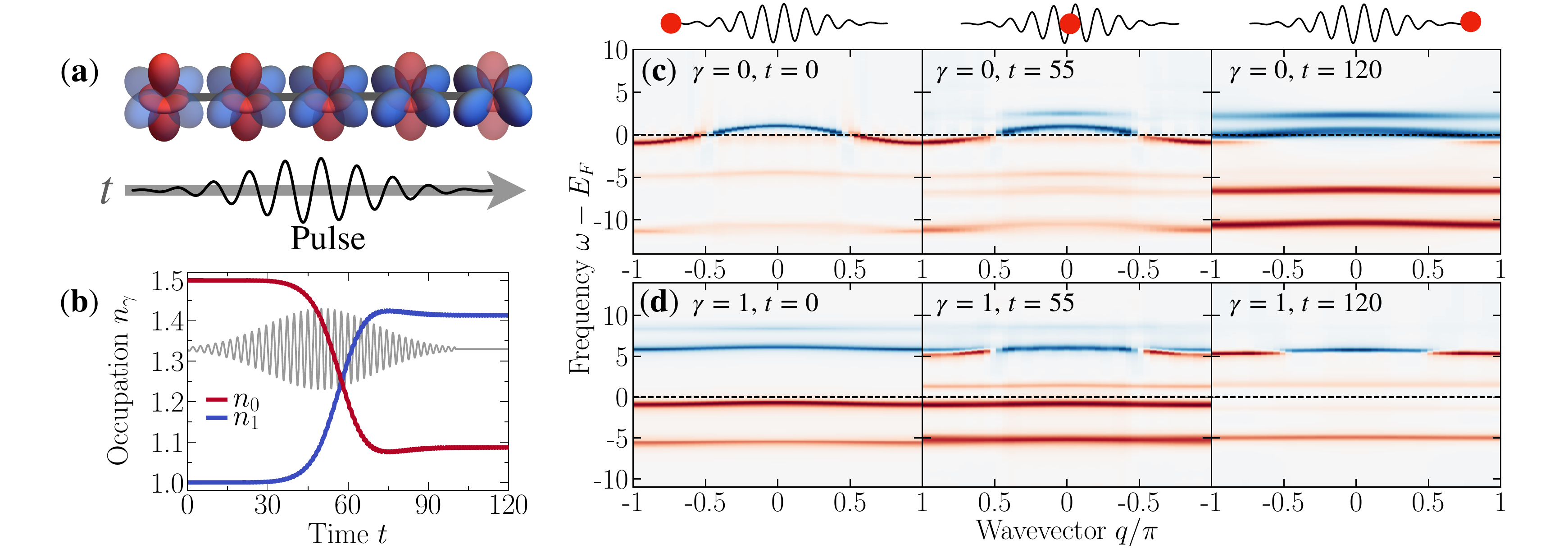}
\caption{(a) Sketch of the photoinduced inversion of the orbital-selective Mott phase in the two-orbital system. The opacity of the orbital lobes indicates the corresponding orbital filling $n_\gamma$. The applied time-dependent pulse changes the orbital fillings during the evolution, leading to a reversal of the occupations. (b) Time evolution of orbital occupancy $n_\gamma$ under a chirped pulse (see the main text for pulse parameters). The gray line shows the pulse profile. (c,d) Orbital-resolved single-particle spectral snapshots $A_\gamma(q,\omega)$ at selected times $t$ before, during, and after the pulse, evaluated with field-free correlation-time propagation. The pulse profile shown above the spectra indicates the time at which each spectral function is evaluated. The upper row [panel (c)] shows the $\gamma = 0$ orbital (itinerant at $t=0$), while the lower row [panel (d)] shows the $\gamma = 1$ orbital (localized at $t=0$). In all panels, red and blue colors denote the electron-removal and electron-addition contributions, respectively. The black dashed line marks the Fermi level of the initial equilibrium state. The calculations were performed for system size $L=24$ at total filling $n=2.5$ (see the main text for the Hamiltonian parameters).}
\label{fig::fig1}
\end{figure*}

{\it Model \textemdash}
We consider a one-dimensional two-orbital Hubbard--Kanamori model previously used to study OSMPs~\cite{Arita2005,Rincon2014-1,Rincon2014-2,Herbrych2019,Herbrych2020-2}. In its spin-rotationally invariant form, the Hamiltonian reads
\begin{equation}
\begin{aligned}
H_0 &=\sum_{\gamma,\gamma',\ell,\sigma}
t_{\gamma\gamma'}
\left(
c^{\dagger}_{\gamma,\ell,\sigma}
c_{\gamma',\ell+1,\sigma}
+\mathrm{H.c.}
\right)
+\Delta\sum_{\ell} n_{1,\ell}
\\
&+U\sum_{\gamma,\ell}
n_{\gamma,\ell,\uparrow}n_{\gamma,\ell,\downarrow}
+\left(U-\frac{5J_{\mathrm{H}}}{2}\right)
\sum_{\ell} n_{0,\ell}n_{1,\ell}
\\
&-2J_{\mathrm{H}}\sum_{\ell}
\mathbf{S}_{0,\ell}\cdot\mathbf{S}_{1,\ell}
+J_{\mathrm{H}}\sum_{\ell}
\left(
P^{\dagger}_{0,\ell}P_{1,\ell}
+\mathrm{H.c.}
\right)\,,
\end{aligned}
\label{eq:H_k}
\end{equation}
where $c^{\dagger}_{\gamma,\ell,\sigma}$ ($c_{\gamma,\ell,\sigma}$) creates (annihilates) an electron with spin $\sigma\in\{\uparrow,\downarrow\}$ in orbital $\gamma\in\{0,1\}$ at site $\ell\in\{1,\ldots,L\}$. Here, $\Delta$ denotes the crystal-field splitting (the onsite-energy offset of the orbital $\gamma=1$), while $n_{\gamma,\ell}=\sum_{\sigma} n_{\gamma,\ell,\sigma}$ is the total electron density in orbital $\gamma$ at site $\ell$. $U$ is the intraorbital Hubbard repulsion, and $J_{\mathrm H}$ is the Hund exchange coupling between the orbital spins $\mathbf S_{\gamma,\ell}$. Finally, $P^{\dagger}_{0,\ell}P_{1,\ell}$ describes interorbital pair hopping, where $P_{\gamma,\ell}=c_{\gamma,\ell,\uparrow}c_{\gamma,\ell,\downarrow}$. We impose open boundary conditions. In the following we use $t_{00}=0.5$, $t_{11}=0.1$, and $t_{01}=t_{10}=0.1$. This choice of parameters is motivated by {\it ab-initio} calculations for low-dimensional multiorbital systems exhibiting an OSMP~\cite{Daghofer2010,Luo2013,Rincon2014-1}. For the results of the main text, we set $\Delta=5.5$. We use $U/W=3$, where $W=2.77$ is the noninteracting bandwidth, and fix $J_{\mathrm H}=U/4$. The total filling is fixed to $n=2.5$ electrons per site. We set $\hbar=1$; energies are quoted in eV and time in units of $\hbar/\mathrm{eV}\simeq0.7$~fs. Pulse frequencies are given in the corresponding units, so a quoted frequency $\omega$ corresponds to photon energy $\hbar\omega$.

{\it Light pulse and time evolution \textemdash}
The coupling to an external time-dependent electric field is introduced through the Peierls substitution~\cite{Aoki2014,Murakami2025}: \mbox{$t_{\gamma\gamma'}\rightarrow t_{\gamma\gamma'}(t)=\exp\left[iA(t)a\right]t_{\gamma\gamma'}$} [and consequently $H_0\rightarrow H(t)$]. Here, $a=1$ is the lattice spacing and $A(t)$ is the vector potential of the applied pump pulse, which is related to the electric field by $E(t)=-\partial_t A(t)$.

Since the two orbitals have different dispersions and bandwidths, a monochromatic pulse addresses only a limited range of states within the excitation-energy window. To excite states over a wider momentum and energy range, we use a linearly chirped pulse whose instantaneous frequency increases from $\omega_{1}$ to $\omega_{2}$~\cite{Werner2019}. The vector potential is chosen as
\begin{equation}
A(t)=a_{0}
\exp\left[
-\frac{(t-t_{P})^{2}}{2\sigma_{P}^{2}}
\right]
\cos\left[\phi(t)\right]\,,
\label{eq::pulse}
\end{equation}
where the phase $\phi(t)$ is defined by the instantaneous frequency,
\begin{equation}
\label{eq::chirped}
\frac{d\phi(t)}{dt}=\omega(t)\,,
\qquad
\omega(t)=\omega_{1}+\alpha(t-t_{i})\,,
\end{equation}
with $\alpha=(\omega_{2}-\omega_{1})/(t_{f}-t_{i})$. Here, $t_{i}=0$ and $t_{f}=2t_P$ specify the interval over which the quoted frequencies are swept. The non-chirped pulse is recovered in the limiting case $\omega_{1}=\omega_{2}=\omega_{P}$. In Eq.~\eqref{eq::pulse}, $a_0$ denotes the pulse amplitude, while the pulse width and center are fixed at $\sigma_P=20$ and $t_P=50$, respectively. Other pulse parameters are presented in the Supplementary Material~\cite{supp}. The nonequilibrium dynamics induced by the pump pulse is calculated using the time-dependent variational principle (TDVP), starting from the ground state $|\text{GS}\rangle$ obtained by the density-matrix renormalization group (DMRG). DMRG and TDVP calculations are performed using the ITensor package~\cite{itensor-1,itensor-2}.

To characterize the evolution of the electronic structure, we evaluate orbital-resolved single-particle spectral snapshots $A_\gamma(q,\omega)$ at selected evolution times $t$~\cite{supp}. The electron-addition and electron-removal correlation functions are evaluated using the time-evolved state $|\psi(t)\rangle$ as the reference state and the field-free Hamiltonian $H_0$ for propagation in the correlation time. Consequently, these spectra characterize the pumped state under field-free propagation.

{\it Photoinduced reversal of orbital selectivity \textemdash}
As illustrated schematically in Fig.~\ref{fig::fig1}(a), the equilibrium ($t=0$) OSMP has an asymmetric distribution of electrons between the orbitals. For the chosen Hamiltonian parameters, the wide $\gamma=0$ orbital has a noninteger filling $n_0=1.5$, while the narrow $\gamma=1$ orbital is half-filled, with $n_1=1.0$ [Fig.~\ref{fig::fig1}(b)]. The Mott-localized character of the orbital $\gamma=1$ and the itinerant character of the orbital $\gamma=0$ are evident in $A_\gamma(q,\omega)$ [left panels of Fig.~\ref{fig::fig1}(c,d)]. Specifically, the orbital $\gamma=0$ exhibits a finite spectral weight at $\omega=E_F$, whereas the orbital $\gamma=1$ displays a Mott gap. The additional bands well below the Fermi level are associated with Hund multiplet excitations~\cite{Bauernfeind2017,Sroda2023}.

Our main result is a substantial orbital population transfer accompanied by a reversal of the character of the orbital-resolved spectral functions, presented in Fig.~\ref{fig::fig1}(c,d). Starting from an OSMP at $t=0$, the system is driven by a chirped pulse with $a_0=2.1$, $\omega_1=2.5$, and $\omega_2=3.1$. Near the pulse center, the occupations become approximately equal, $n_0\approx n_1\approx1.25$ [Fig.~\ref{fig::fig1}(b)]. In both orbitals, the electron-addition and electron-removal features approach one another near the initial $E_F$ and show appreciable momentum dependence [middle panels of Fig.~\ref{fig::fig1}(c,d)]. After the pulse, the occupations are nearly reversed, with $(n_0,n_1)\approx(1.08,1.42)$, corresponding to a transfer of approximately $0.42$ electrons per site. Orbital $\gamma=0$ is now close to half filling, and its prominent addition and removal features become separated, consistent with increased Mott localization. In contrast, the corresponding features of the orbital $\gamma=1$ move closer together and resemble those of the itinerant (metallic) orbital with the new (effective) Fermi level $\widetilde{E}_F> E_F$. These results show that the localized and itinerant characters of the orbitals are interchanged in the light-induced nonequilibrium state.

\begin{figure}[!htb]
\includegraphics[width=\columnwidth]{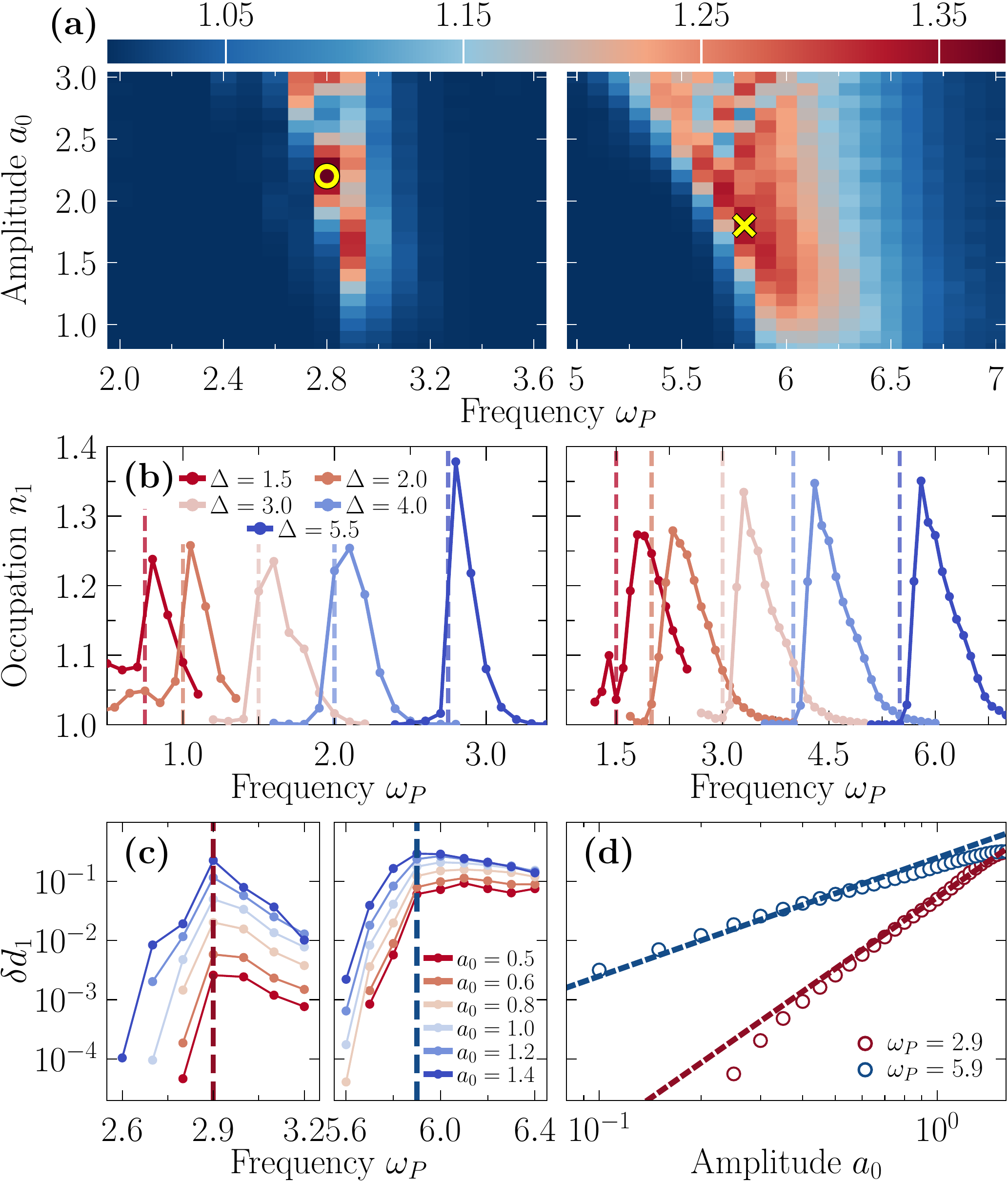}
\caption{(a) Post-pulse ($t=120$) occupation of the initially localized orbital $n_1$ vs pulse amplitude $a_0$ and frequency $\omega_P$. The largest post-pulse occupations are obtained for $a_0=2.2$,~$\omega_P=2.8$ (yellow circle) and $a_0=1.8$,~$\omega_{P}=5.8$ (yellow cross). (b) Post-pulse ($t=120$) occupations $n_1$ for various crystal-field splittings $\Delta$. The left and right panels show results at the lower and higher driving frequencies, $\omega_P=2.8$ and $\omega_P=5.8$, respectively. (c) Post-pulse excess doublon density, $\delta d_1=d_1(a_0)-d_1(a_0=0)$, as a function of $\omega_P$ for several pulse amplitudes $a_0$. The red and blue dashed lines indicate the positions of the two resonances in the low-amplitude regime. (d) Amplitude dependence of $\delta d_1$ at $\omega_P=2.9$ and $5.9$. Dashed lines show power-law fits, yielding $\delta d_1\propto a_0^4$ for $\omega_P=2.9$ and $\delta d_1\propto a_0^2$ for $\omega_P=5.9$. The system size in all panels is $L=16$.}
\label{fig::fig2}
\end{figure}

{\it Pulse parameters and excitation channels \textemdash} 
To quantify the photoinduced population transfer, we focus on the occupation $n_1$ of the initially localized orbital $\gamma=1$, which is half-filled ($n_1=1$) in equilibrium. Since the total filling is conserved, an increase in $n_1$ measures the net transfer from orbital $\gamma=0$ to orbital $\gamma=1$. In the present simulations, the transfer is accompanied by increased double occupancy \mbox{$d_1=(1/L)\sum_\ell n_{1,\ell,\uparrow}n_{1,\ell,\downarrow}$} in the orbital $\gamma=1$, as also indicated by the multiplet decomposition~\cite{supp}. In the following, we use $n_1$ and $d_1$ as a transfer diagnostic, while the analysis of the orbital localization is additionally based on the $A(q,\omega)$ spectra.

To identify the optimal driving frequencies, we first scan non-chirped pulses, $\omega_1=\omega_2=\omega_P$, to identify the frequencies and amplitudes that maximize interorbital population transfer. Figure~\ref{fig::fig2}(a) shows the post-pulse occupation ($t=120$) of the orbital $\gamma=1$, $n_1$, as a function of pulse amplitude $a_0$ and frequency $\omega_P$. Two regions of enhanced population transfer are clearly visible around $a_0=2.2$, $\omega_{P}=2.8$ and around $a_0=1.8$, $\omega_{P}=5.8$, indicated by the yellow circle and cross, respectively. Note that these frequencies are close to $\omega_P=\Delta/2$ and $\omega_P=\Delta$ (with $\Delta=5.5$ used here), indicating two distinct excitation channels. In Fig.~\ref{fig::fig2}(b), we show that this qualitative behavior is found for all considered $\Delta$ (provided that the initial state is in the OSMP regime).

Furthermore, we analyze the photoinduced change in the doublon density, \mbox{$\delta d_1=d_1(a_0)-d_1(a_0=0)$}, which removes the equilibrium contribution. In the perturbative regime, the excitation rate of an $m$-photon process scales as $F_0^{2m}$, where $F_0$ is the electric-field amplitude~\cite{Oka2012}. Since $F_0\propto\omega_P a_0$ at a fixed frequency, one expects~\mbox{$\delta d_1\propto a_0^{2m}$}. Figure~\ref{fig::fig2}(c) shows $\delta d_1$ as a function of $\omega_P$ for several pulse amplitudes. Although the strongest population transfer in Fig.~\ref{fig::fig2}(a) occurs near $\omega_P=2.8$, this maximum lies in the strong-field regime. At smaller amplitudes, relevant for the perturbative analysis, the lower-frequency resonance shifts slightly toward $\omega_P\simeq2.9$, as indicated by the red dashed line in Fig.~\ref{fig::fig2}(c). The higher-frequency resonance is wider and shows a much weaker amplitude-dependent shift; therefore, we use $\omega_P=5.9$ as a representative frequency for this channel. As shown in Fig.~\ref{fig::fig2}(d), the weak-field response at $\omega_P=2.9$ follows approximately $\delta d_1\propto a_0^4$, consistent with a two-photon process, while at $\omega_P=5.9$ it follows $\delta d_1\propto a_0^2$ over a broader amplitude range, consistent with a direct one-photon process.

\begin{figure}[!ht]
\includegraphics[width=\columnwidth]{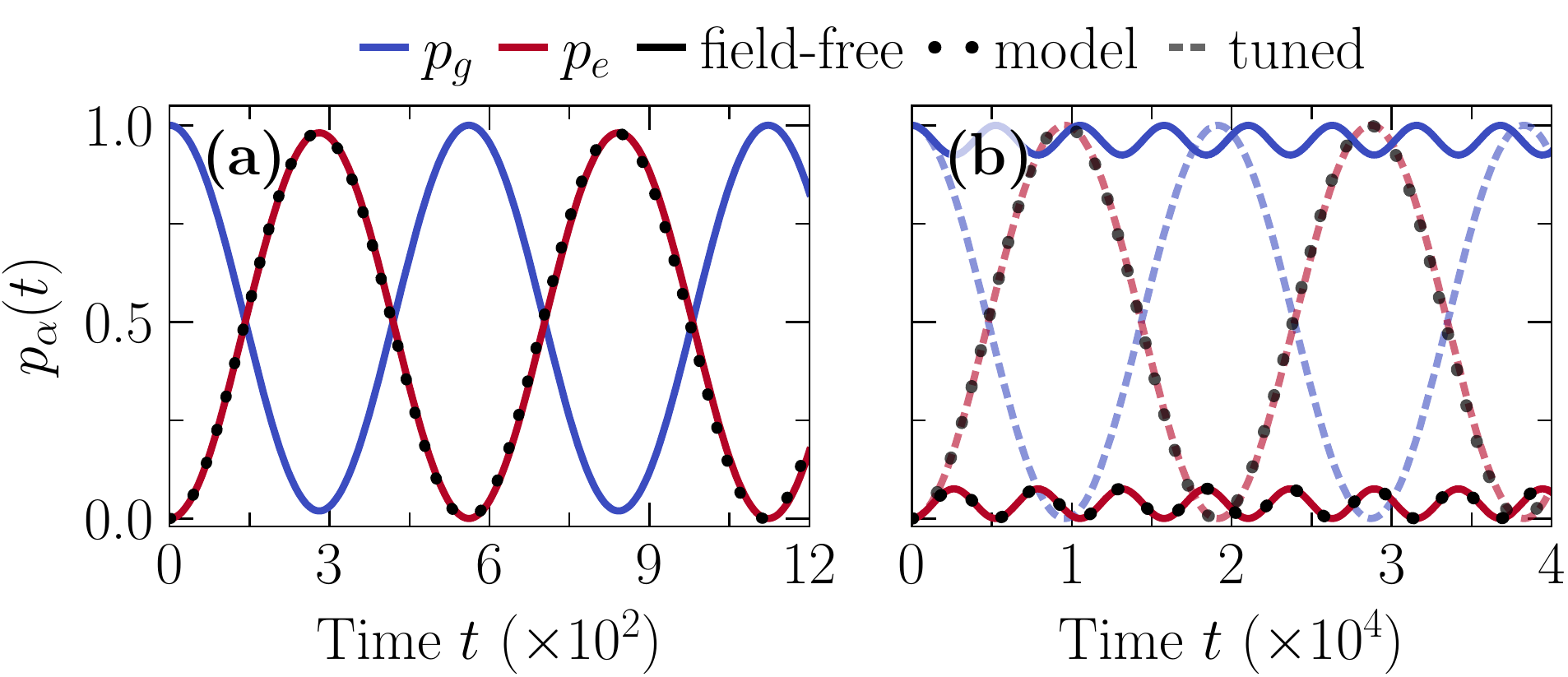}
\caption{(a) One-photon population dynamics of the $L=2$ system at the field-free resonance, $\omega_P=6.10365$. (b) Two-photon dynamics at the field-free resonance, $\omega_P=2.95169$ (solid curves) and drive-shifted resonance, $\omega_P=2.95112$ (dashed curves). Both panels show the exact evolution under $A(t)=a_0\cos(\omega_P t)$ with $a_0=0.1$. Blue and red curves plot $p_\alpha(t)=|\langle\alpha|\psi(t)\rangle|^2$ for the field-free ground state ($\alpha=g$) and the target excited state ($\alpha=e$), respectively. 
Black dots show $p_e(t)$ predicted by the effective two-level model (see Eq.~\eqref{eq:S_population} of \cite{supp}), with oscillation frequency $\Omega_{\mathrm{osc}}^{(n)}$ and maximum population $p_{e,\max}^{(n)}$ given in Eq.~\eqref{eq:main_transfer_contrast}.
}
\label{fig::fig3}
\end{figure}

Beyond identifying the photon order, we ask how the two channels differ in their coherent population-transfer dynamics. Our analysis indicates that the higher-frequency pulse produces more pronounced Rabi-like oscillations, whereas the lower-frequency pulse leaves a larger post-pulse occupation (see Fig.~\ref{fig::figS2} of~\cite{supp}). A minimal microscopic description follows from exact diagonalization of the two-site system under a constant-amplitude drive, $A(t)=a_0\cos(\omega_Pt)$. Near each resonance, such a weak-drive two-level approximation gives an effective coupling $g_n$ and detuning $\delta_n=E_e-E_g-n\omega_P+\Delta E_e-\Delta E_g$, where $n=1,2$ is the order of the excitation process and $\Delta E_{g,e}$ are the drive-induced energy shifts \cite{PhysRev.138.B979,Guerin2003}. With $\Omega^{(n)}=2|g_n|$, the maximum excited-state population and the oscillation frequency are given by
\begin{equation}
p_{e,\max}^{(n)}=\frac{(\Omega^{(n)})^2}{(\Omega^{(n)})^2+\delta_n^2},
\qquad
\Omega_{\mathrm{osc}}^{(n)}=\sqrt{(\Omega^{(n)})^2+\delta_n^2}.
\label{eq:main_transfer_contrast}
\end{equation}
Here $g_1\propto a_0$, while $g_2$ and the drive-induced energy shifts scale as $a_0^2$. At the field-free resonance, $\omega_P=(E_e-E_g)/n$, where the pump frequency is set by the undriven energy gap, the ratio $|\delta_1|/\Omega^{(1)}$ vanishes linearly with $a_0$, while $|\delta_2|/\Omega^{(2)}$ can remain finite as $a_0\to 0$. Using the two-site estimates at $a_0=0.1$, these ratios are approximately $0.14$ and $3.5$, corresponding to maximum populations of approximately $0.98$ and $0.075$, respectively. Consequently, when the pump frequency is set by the field-free gap, $\omega_P=(E_e-E_g)/n$, the one-photon process [see Fig.~\ref{fig::fig3}(a)] shows nearly complete population exchange, while two-photon transfer is strongly suppressed [see Fig.~\ref{fig::fig3}(b)]. Adjusting $\omega_P$ such that $\delta_2\simeq0$ restores nearly complete two-photon oscillations~\cite{supp}, as shown by the dashed curves in Fig.~\ref{fig::fig3}(b).

At the respective shifted resonances and the same weak-drive amplitude, the smaller two-photon Rabi frequency places both excitation and population return on a longer time scale. This offers a qualitative interpretation of the different transient dynamics, but does not by itself determine which finite pulse leaves a larger transferred population: this also depends on the envelope, duration, and detuning. Moreover, the weak-drive two-site description is not quantitatively predictive for the strong pulses used for orbital inversion. In larger systems, the distribution of transition energies associated with orbital dispersions additionally limits transfer by a monochromatic pulse. Consequently, chirping around the lower-frequency resonance improves transfer by addressing a broader excitation window~\cite{supp}.

{\it Discussion \& Conclusions \textemdash}
We showed that a tailored light pulse can invert an orbital-selective Mott phase. In MPS simulations of a one-dimensional two-orbital Hubbard-Kanamori model, a chirped pump pulse transfers charge from the initially itinerant orbital $\gamma=0$ to the Mott-localized orbital $\gamma=1$, nearly reversing the equilibrium occupations from $(n_0,n_1)=(1.5,1.0)$ to $(1.08,1.42)$. The single-particle spectral snapshots at various times show that this redistribution does not merely metallize both orbitals; instead, orbital selectivity re-emerges after the pulse with the localized and itinerant roles interchanged. The transfer proceeds via two excitation channels - a one-photon process at $\omega_P\simeq\Delta$ and a two-photon process at $\omega_P\simeq\Delta/2$ - and is maximized by a linearly chirped pulse tuned to the lower-frequency resonance.

The inverted OSMP is remarkably long-lived: after the pulse, $n_1$ reaches a plateau that persists far beyond the drive~\cite{supp}. This stability is not enforced by any conservation law -  interorbital and pair hopping do allow charge to flow back - but it reflects a relaxation bottleneck set by the large gap-to-bandwidth ratio of the OSMP~\cite{Sensarma2010,Eckstein2011,Hofmann2012}. Recovering the initial state requires the decay of doublon-like configurations in orbital $\gamma=1$, the back-transfer of electrons to orbital $\gamma=0$, and the redistribution of the absorbed energy among the remaining degrees of freedom, all of which are higher-order processes much slower than the photoinduced transfer. The effect survives in the thermodynamic limit $n_1(t\gg t_P,L\rightarrow\infty)\simeq1.42$, and persists at least up to $t\sim300$~\cite{supp}. In real materials, coupling with phonons~\cite{Lenarcic2015,Murakami2025} and magnons \cite{Avigo2016} will open additional relaxation channels and ultimately limit this lifetime.

The OSMP switching relies on a few generic ingredients: a half-filled orbital stabilized as a Mott insulator, a metallic orbital at noninteger filling, crystal-field splitting, and finite interorbital hopping, which enables light-driven charge transfer between the orbitals through the Peierls coupling. The large $\Delta$ used here clearly separates the two channels, but the resonance conditions scale with $\Delta$, so realistic splittings of $0.1$-$1$~eV place them in the infrared range accessible to current strong-field sources. Natural candidates are the quasi-one-dimensional iron-based ladders BaFe$_2$Se$_3$ and BaFe$_2$S$_3$~\cite{Caron2012,Takahashi2015,Yamauchi2015,Mourigal2015}, for which the present model was originally constructed~\cite{Rincon2014-1,Herbrych2021}, and other iron chalcogenides, where angle-resolved photoemission (ARPES) reveals selective localization of the $d_{xy}$ states~\cite{Yi2013}. A further candidate is Ca$_{2-x}$Sr$_x$RuO$_4$, the original setting of the OSMP~\cite{Anisimov2002}, for which ARPES provides evidence of selective Mott localization~\cite{Neupane2009}, although this interpretation remains debated~\cite{Wang2004,Shimoyamada2009}; its parent compound Ca$_2$RuO$_4$ realizes a closely related configuration, with a fully occupied $d_{xy}$ orbital coexisting with Mott-localized $d_{xz}/d_{yz}$ states~\cite{Sutter2017}. Furthermore, layered nickelates also exhibit pronounced orbital-dependent correlations~\cite{Liao2023,Cao2024,Yang2024,Liao2026,Li2017}. For La$_3$Ni$_2$O$_7$, the nominal $e_g$ filling of $1.5$ is the particle-hole analogue to the case $n=2.5$ studied here; La$_4$Ni$_3$O$_{10}$ instead has a nominal filling $4/3$, which can be related to $8/3$. Theory and experiment consistently report strongly orbital-selective correlations in these compounds, with nearly localized $d_{z^2}$ and itinerant $d_{x^2-y^2}$ electrons, placing them in proximity to an OSMP~\cite{Liao2023,Cao2024,Yang2024,Liao2026,Li2017}.

Finally, when the relevant orbitals contribute differently to currents along different directions, redistribution among them could alter the polarization-dependent transient optical conductivity. In nickelates, $d_{z^2}$ is predominantly coupled along the $c$ axis through the apical oxygen, while $d_{x^2-y^2}$ disperses in the plane~\cite{Cao2024,Liao2026}; similar spatial differentiation arises between the $a_{1g}$ and $e_g'$ states of quasi-one-dimensional BaVS$_3$~\cite{Lechermann2005} and between $d_{xz}$ and $d_{yz}$ in orthorhombically distorted or uniaxially strained ruthenates. In these systems, the photoinduced inversion should manifest as a light-driven switching of the transport anisotropy, detectable by polarization-resolved time-resolved terahertz or optical conductivity measurements, without orbital-resolved spectroscopy.

{\it Acknowledgments}
M.L. and J.H. acknowledge the support of the National Science Centre (NCN), Poland, through the project 2023/50/E/ST3/00033. J.H. acknowledges support from the Invitational Fellowships of the Japan Society for the Promotion of Science (JSPS), as well as the hospitality of Tokyo University of Science, where part of this work was conducted. The calculations were performed using resources provided by the Wroc{\l}aw Centre for Networking and Supercomputing (\url{http://wcss.pl}). Data supporting the findings of this article are openly available~\cite{opendata}.

\bibliography{inverseosmp.bib}
\phantom{a}
\newpage
\clearpage

\setcounter{figure}{0}
\setcounter{equation}{0}
\setcounter{table}{0}

\renewcommand{\thetable}{S\arabic{table}}
\renewcommand{\thefigure}{S\arabic{figure}}
\renewcommand{\theequation}{S\arabic{equation}}
\renewcommand{\thepage}{S\arabic{page}}

\setcounter{section}{0}
\setcounter{secnumdepth}{1}
\renewcommand{\thesection}{S\arabic{section}}

\onecolumngrid

\begin{center}
{\large \bf Supplemental Material:\\
Photoinduced switching of orbital-selective Mott states}\\
\vspace{0.3cm}
M. Lisiecki, P. Werner, and J. Herbrych
\setcounter{page}{1}
\end{center}
\vspace{0.6cm}
\twocolumngrid
\label{pagesupp}

\section{Single-particle spectral function} \label{app:a}
 Spectral snapshots are constructed from site-resolved correlation functions with a fixed reference state $|\psi(t)\rangle$. For a given orbital $\gamma$ and spin $\sigma$, we separate the correlation function into the greater, hole-like contribution and the lesser, electron-like contribution, defined as
 \begin{equation}
\begin{aligned}
G^{\mathrm{h}}_{\gamma,\sigma}(\ell,\omega)
&= -i\int_0^{T_{\max}} d\tau\,
   e^{i(\omega+i\eta)\tau} \\
&\quad\times
\langle\psi(t)|c_{\gamma,\ell,\sigma}(\tau)
c_{\gamma,L/2,\sigma}^{\dagger}|\psi(t)\rangle,\\[2pt]
G^{\mathrm{e}}_{\gamma,\sigma}(\ell,\omega)
&= -i\int_0^{T_{\max}} d\tau\,
   e^{i(\omega+i\eta)\tau} \\
&\quad\times
\langle\psi(t)|c_{\gamma,\ell,\sigma}^{\dagger}
c_{\gamma,L/2,\sigma}(\tau)|\psi(t)\rangle.
\end{aligned}
\end{equation}
Here, $\eta=0.2$ is the exponential broadening and $T_\text{max}$ is the maximum relative evolution time $\tau$, sampled with a time step $\Delta\tau=0.01$. The operators evolve according to $c_{\gamma,\ell, \sigma}(\tau)=U^\dagger(\tau) c_{\gamma,\ell, \sigma} U(\tau)$ with $U(\tau)=\exp(-iH_0\tau)$ and the time-evolved state is given by \mbox{$|\psi(t)\rangle=\mathcal{T}\exp[-i\int_{0}^{t}\mathrm{d}t^\prime\,H(t^\prime)]|\text{GS}\rangle$}, where $\mathcal{T}$ denotes the time-ordering operator. The corresponding spectral functions are obtained by taking the imaginary parts of these Green's functions
\begin{equation}
\begin{aligned}
A^{\mathrm{h}}_{\gamma,\sigma}(\ell,\omega)
&= -\frac{1}{\pi}\mathrm{Im}\,
G^{\mathrm{h}}_{\gamma,\sigma}(\ell,\omega)\,, \\
A^{\mathrm{e}}_{\gamma,\sigma}(\ell,\omega)
&= -\frac{1}{\pi}\mathrm{Im}\,
G^{\mathrm{e}}_{\gamma,\sigma}(\ell,\omega)\,.
\end{aligned}
\end{equation}
Finally, the single-particle momentum-resolved spectral function is obtained by Fourier transforming the site-resolved spectral functions with respect to the center site $L/2$,
\begin{equation}
A_{\gamma}(q,\omega)
=
\sum_{\ell, \sigma} e^{-iq(\ell-L/2)}
\left[
A^{\mathrm{h}}_{\gamma,\sigma}(\ell,\omega)
+
A^{\mathrm{e}}_{\gamma,\sigma}(\ell,\omega)
\right]\,.
\label{eq:SSF}
\end{equation}
The corresponding orbital-resolved density-of-states ($\mathrm{DOS}(\omega)$) is obtained by summing the spectral function over discrete momenta $A_\gamma(\omega)=(1/L)\sum_{q} A_{\gamma}(q,\omega)$. The reference point \(L/2\) is chosen according to the central-site trick. This approximation reduces the Fourier transform in Eq.~\eqref{eq:SSF} from a costly double sum over pairs of sites to a single sum over distances measured from the center of the lattice. The procedure is exact for periodic boundary conditions, where all sites are equivalent, and remains accurate for open boundary conditions, for which the central site is the least affected by boundary effects.

\begin{figure}[!htb]
\includegraphics[width=\columnwidth]{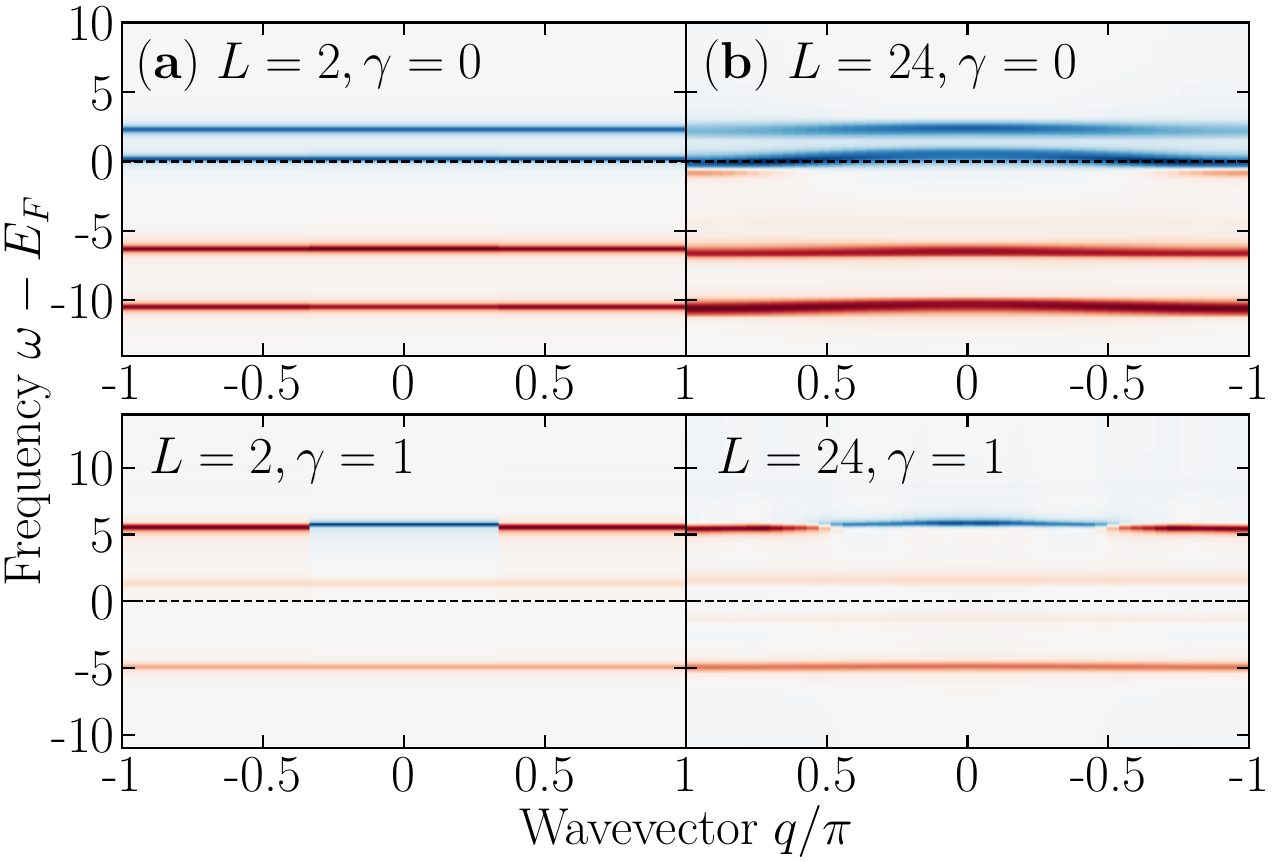}
\caption{Post-pulse orbital-resolved single-particle spectral functions $A_{\gamma}(q,\omega)$ at $t=120$. The left column shows results for system size $L=2$ after a non-chirped pulse with $a_{0}=2.2$ and $\omega_{P}=2.8$, while the right column shows results for $L=24$ after a linearly chirped pulse with $a_{0}=2.1$, $\omega_{1}=2.5$, and $\omega_{2}=3.1$. The upper and lower rows correspond to orbitals $\gamma=0$ and $\gamma=1$, respectively. Red and blue denote the electron- and hole-like contributions, respectively, and the black dashed line marks the Fermi level of the initial equilibrium state.}
\label{fig::figS1}
\end{figure}

Figure~\ref{fig::figS1} compares the post-pulse orbital-resolved spectra obtained with two representative driving protocols. For $L=2$, the non-chirped pulse with $a_0=2.2$ and $\omega_P=2.8$ addresses the lower-frequency two-photon channel and drives a complete population reversal, accompanied by an exact exchange of the electron- and hole-like features between the orbitals. For $L=24$, we used a chirped pulse around the two-photon resonance ($a_0=2.1$, $\omega_1=2.5$, and $\omega_2=3.1$). In this case, the broader range of momentum-resolved transitions and the partial charge transfer result in an only approximate spectral inversion. Nevertheless, the initially itinerant $\gamma=0$ orbital develops separate electron- and hole-like features consistent with localization, while the $\gamma=1$ orbital acquires a more dispersive, approximately half-filled spectrum, demonstrating that the orbital-selective character is inverted.

\section{Pulse optimization} \label{app:b}

\begin{figure}[!htb]
\includegraphics[width=\columnwidth]{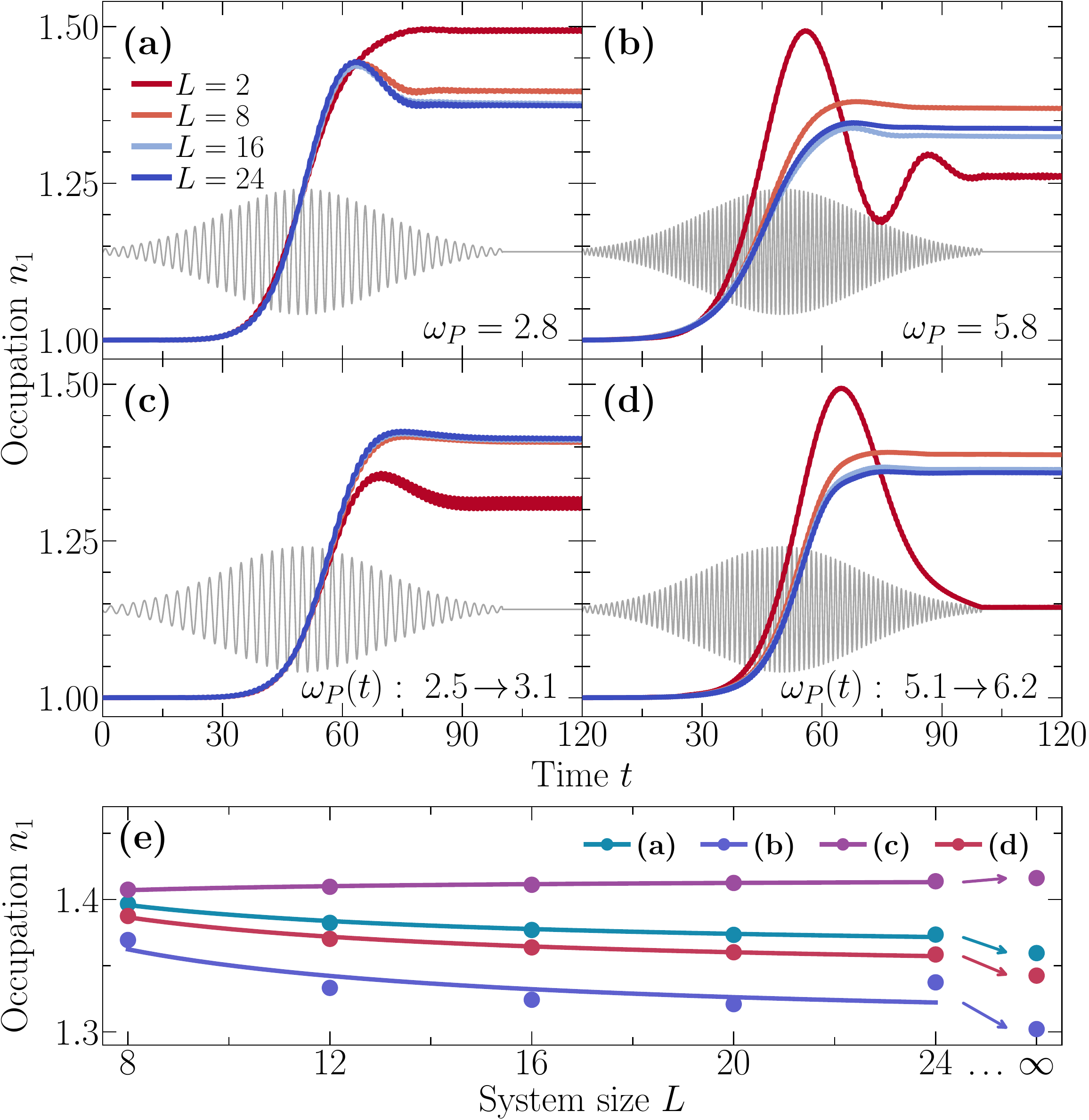}
\caption{Time evolution of the orbital occupation $n_1$ for different system sizes under four driving protocols: (a) non-chirped pulse with $a_0=2.2$ and $\omega_P=2.8$, (b) non-chirped pulse with $a_0=1.8$ and $\omega_P=5.8$, (c) chirped pulse with $a_0=2.1$, whose frequency increases linearly from $\omega_1=2.5$ to $\omega_2=3.1$, (d) chirped pulse with $a_0=1.6$, whose frequency increases linearly from $\omega_1=5.1$ to $\omega_2=6.2$. In panels (a--d), results are shown for $L=2$, $8$, $16$, and $24$, while the gray curves indicate the corresponding pulse profiles. (e) Finite-size scaling of the post-pulse  occupation $n_1(t=120)$ for the four driving protocols shown in panels (a--d). The rightmost points indicate the thermodynamic-limit values $n_1 (L\rightarrow\infty)\simeq1.36$, $1.30$, $1.42$, and $1.34$ for panels (a), (b), (c), and (d), respectively, obtained from the $1/L$ fit.}
\label{fig::figS2}
\end{figure}

Figure~\ref{fig::figS2}(a--d) shows the time evolution of $n_1$ for different system sizes and for the driving protocols considered in the main text. In all cases, $n_1$ increases during the pulse and approaches a long-lived post-pulse plateau. A particularly notable finite-size effect appears for the non-chirped two-photon excitation in Fig.~\ref{fig::figS2}(a). For $L=2$, the pulse drives the system to complete orbital inversion. However, with increasing system size, the post-pulse value of $n_1$ decreases. As the system size increases, a larger number of momentum states contributes to the dynamics, giving rise to a broader set of transition energies. Since the two orbitals have different bandwidths, a single-frequency pulse cannot efficiently excite all relevant transitions, thereby reducing the population-transfer efficiency. For the direct one-photon excitation shown in Fig.~\ref{fig::figS2}(b), finite-size effects are even more apparent in the transient dynamics.

In Figs.~\ref{fig::figS2}(c,d) we present the results for the linearly chirped pulses. Since the instantaneous frequency varies over a finite interval, the pulse addresses a broader range of transition energies and therefore better accommodates the different orbital dispersions. In particular, for the chirped pulse associated with the two-photon channel, Fig.~\ref{fig::figS2}(c), the post-pulse occupation $n_1$ shows only a weak dependence on system size. Chirping the higher-frequency one-photon channel, Fig.~\ref{fig::figS2}(d), also enhances the population transfer compared with the corresponding non-chirped drive, although the final occupation remains lower than for the chirped two-photon channel.

To quantify the behavior in the thermodynamic limit, Fig.~\ref{fig::figS2}(e) shows the post-pulse occupancy $n_1(t=120)$ as a function of $L$. Data for $L=8$, $12$, $16$, $20$, and $24$ are fitted to $n_1(L)=n_1(L \rightarrow \infty)+A/L$. The extrapolated values are $n_1(L \rightarrow \infty)\simeq1.36$, $1.30$, $1.42$ and $1.34$ for the protocols shown in panels (a), (b), (c) and (d), respectively. Thus, the enhanced post-pulse occupation persists in the thermodynamic limit for both excitation channels, with the largest inversion obtained for the chirped pulse around the lower-frequency two-photon resonance.

\begin{figure}[!t]
\includegraphics[width=\columnwidth]{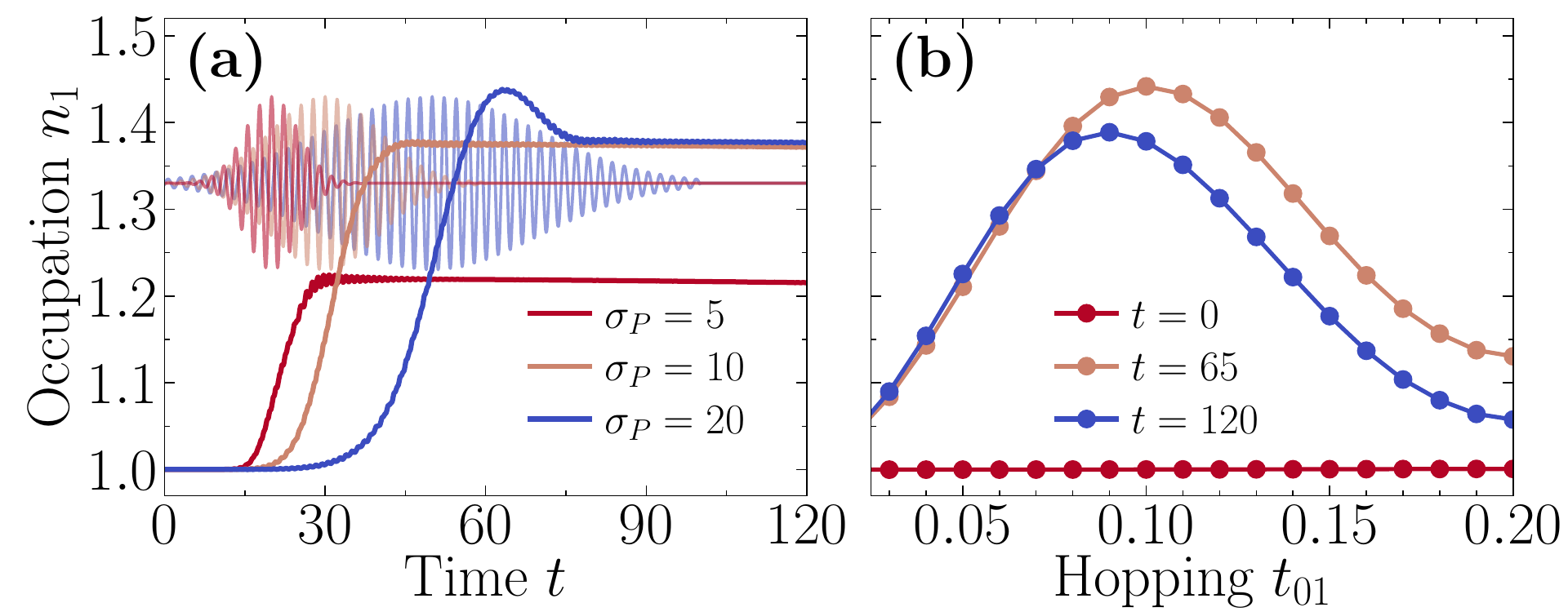}
\caption{(a) Time evolution of $n_1$ under pulses with fixed frequency $\omega_P=2.8$ for various pulse widths $\sigma_P=5$, $10$, and $20$, using amplitudes $a_0=3.0$, $2.5$, and $2.2$, respectively. The corresponding pulse profiles are also presented. (b) Occupation $n_1$ as a function of the interorbital hopping $t_{01}=t_{10}$ before ($t=0$), during ($t=65$), and after ($t=120$) non-chirped pulses with $a_0=2.2$, $\omega_P=2.8$, $\sigma_P=20$, and $t_P=50$. The system size for both panels is $L=16$.}
\label{fig::figS3}
\end{figure}

Another relevant control parameter is the pulse width. Figure~\ref{fig::figS3}(a) compares the time evolution of $n_1$ for $\sigma_P=5$, $10$ and $20$, using progressively smaller amplitudes $a_0=3.0$, $2.5$ and $2.2$, respectively. Our results indicate that longer pulses produce a larger post-pulse occupation of the orbital $\gamma=1$ despite the reduced peak amplitude, indicating a more efficient photodoping and inversion of the OSMP.

Figure~\ref{fig::figS3}(b) shows the nonmonotonic dependence of the post-pulse occupation $n_1$ on the interorbital hopping $t_{01}=t_{10}=t_{\gamma\ne\gamma^\prime}$. The results indicate that moderate interorbital hopping can enhance the occupation transfer associated with photoinduced inverted OSMP. This behavior reflects two competing effects: on the one hand, larger interorbital hopping facilitates the transfer of electron density between orbitals, increasing the efficiency of the inversion; on the other hand, it also accelerates the relaxation of the transferred charge after the pulse. Indeed, for $t_{01}=t_{10}>t_{11}$, $n_1$ decreases more between $t=65$ and $t=120$, consistent with faster post-pulse relaxation. Note also that increasing $t_{\gamma\ne\gamma^\prime}$ brings the equilibrium system closer to the OSMP boundary. The value $t_{01}=0.1$ used in the main text yields a post-pulse occupation close to the maximum within the range considered.

\begin{figure}[!t]
\centering
\includegraphics[width=\columnwidth]{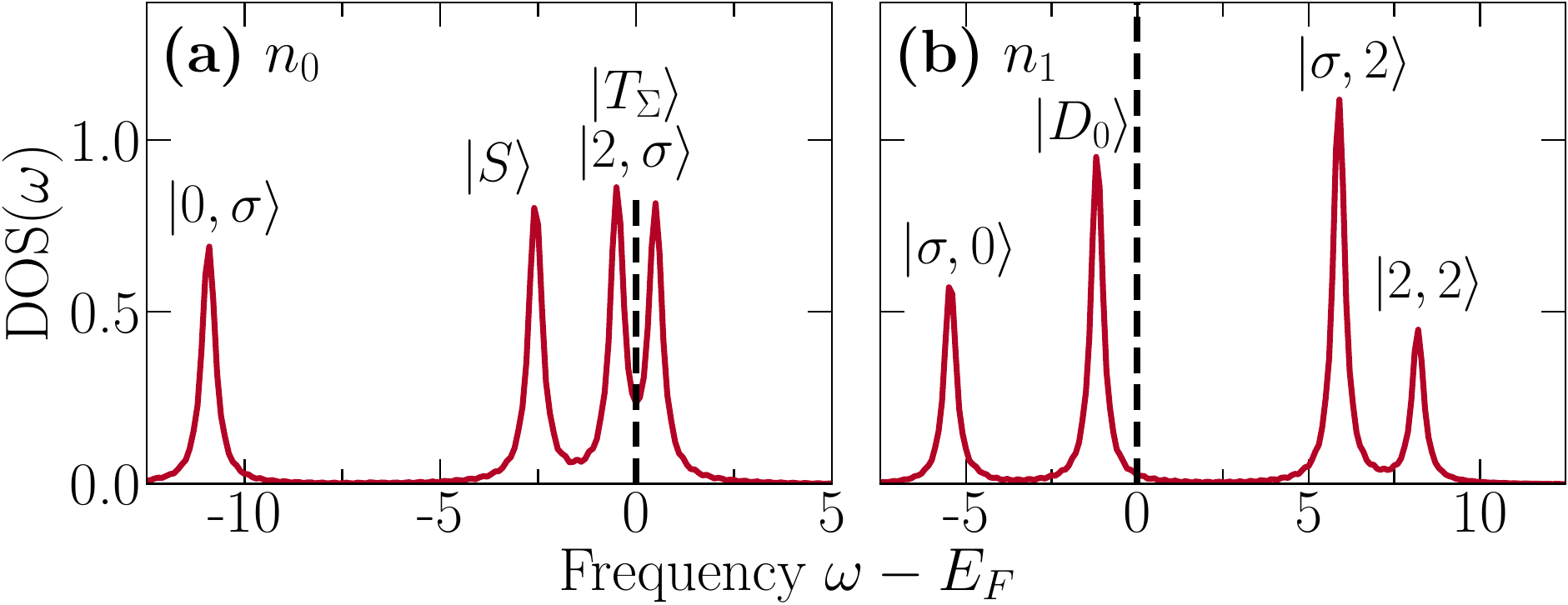}
\caption{Equilibrium orbital-resolved density-of-states $\mathrm{DOS}(\omega)$. Panel~(a) depicts initially itinerant orbital $\gamma=0$, while panel (b) initially Mott-localized orbital $\gamma=1$. The vertical dashed line marks the Fermi level. Results are shown for $L=2$.}
\label{fig::figS4}
\end{figure}

\section{Multiplet decomposition} \label{sup::atomic}

To identify the local configurations involved in the photoinduced charge transfer, we analyze the dynamics in terms of the eigenstates of the atomic-limit Hamiltonian. The resulting local eigenstates and their excitation energies for the parameters used in this work are summarized in Table~\ref{tab:two_orbital_eigenstates}. Finite hopping hybridizes these local configurations and broadens the corresponding atomic excitations into dispersive bands. Nevertheless, the atomic eigenstates provide a useful local basis for resolving the dominant character of the many-body state and of the photoinduced excitations. In Fig.~\ref{fig::figS4}, we show the orbital-resolved density-of-states, with the main spectral features labeled according to the corresponding atomic-limit transitions listed in Table~\ref{tab:two_orbital_eigenstates}.

\begin{figure}[!t]
\includegraphics[width=1.0\columnwidth]{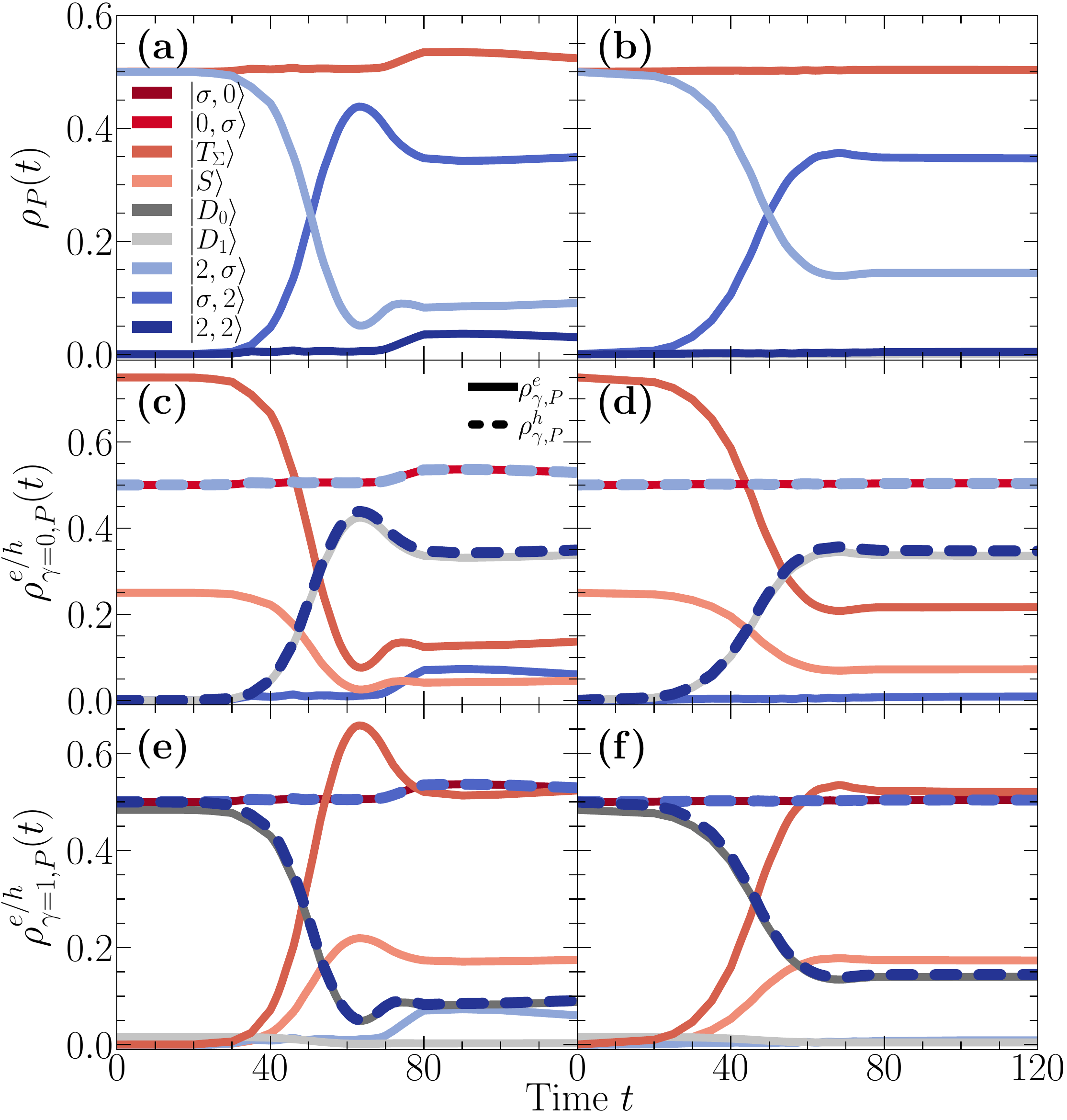}
\caption{Time evolution of the local multiplet decomposition and the multiplet-resolved single-particle excitation weights. The left column, panels (a,c,e), corresponds to the two-photon channel with $\omega_P=2.8$ and $a_0=2.2$, while the right column, panels (b,d,f), corresponds to the direct one-photon channel with $\omega_P=5.8$ and $a_0=1.8$. Panels (a,b) show the decomposition of the time-evolved state into local atomic multiplets, quantified by $\rho_P(t)$. Panels (c,d) and (e,f) show the multiplet-resolved electron and hole excitation weights for orbitals $\gamma=0$ and $\gamma=1$, respectively, with solid and dashed lines denoting the electron and hole contributions. Results are shown for $L=16$ and $\Delta=5.5$.}
\label{fig::figS5}
\end{figure}

In order to obtain insight into the time dependence of the excitations, we introduce local projectors $P_\ell$ onto the atomic multiplets. For example, the projector onto the threefold triplet sector is given by
\begin{equation}
P_{T,\ell}=|T_1\rangle\langle T_1|+|T_0\rangle\langle T_0|+|T_{-1}\rangle\langle T_{-1}|\,,
\end{equation}
where \mbox{$|T_1\rangle=|\uparrow,\uparrow\rangle$}, \mbox{$|T_0\rangle=\frac{1}{\sqrt{2}}(|\uparrow,\downarrow\rangle+|\downarrow,\uparrow\rangle)$}, and \mbox{$|T_{-1}\rangle=|\downarrow,\downarrow\rangle$}. Analogous projectors are constructed for the singlet, doublon, and three-electron sectors. 

\begin{table*}[!htb]
\centering
\renewcommand{\arraystretch}{1.45}
\setlength{\tabcolsep}{5pt}
\caption{Eigenstates and eigenvalues of the atomic-limit two-orbital Hamiltonian. The states are written in the basis $\left|\gamma_0,\gamma_1\right\rangle$. Here we use $n=2.5$, $\sigma=\{\uparrow,\downarrow\}$, $\Sigma=\{-1,0,1\}$, $J_\mathrm{H}=U/4$, and $\mu=2U-2J_\mathrm{H}$. The last column represents the excitation energy relative to the assumed ground state $|\mathrm{GS}\rangle\propto |T_\Sigma\rangle + |2,\sigma\rangle$, i.e., $(E-\mu N)-(E_{\mathrm{gs}}-\mu N_{\mathrm{gs}}$).} 
\label{tab:two_orbital_eigenstates}
\begin{tabular}{|c|c|c|c|c|}
\hline
$N$ &
Degeneracy &
Label &
Atomic limit eigenstate&
Excitation energy w.r.t. GS\\
\hline

\multirow{2}{*}{$1$}
& $2$ &
$\left|\sigma,0\right\rangle$ &
$\left|\uparrow,0\right\rangle,\left|\downarrow,0\right\rangle$ &
$U+J_\mathrm{H}-\Delta$ \\
\cline{2-5}

& $2$ &
$\left|0,\sigma\right\rangle$ &
$\left|0,\uparrow\right\rangle,\left|0,\downarrow\right\rangle$ &
$U+J_\mathrm{H}$ \\
\hline

\multirow{4}{*}{$2$}
& $3$ &
$\left|T_{\Sigma}\right\rangle$ &
$\left|\uparrow,\uparrow\right\rangle$,
$\frac{1}{\sqrt{2}}
\left(
\left|\uparrow,\downarrow\right\rangle+
\left|\downarrow,\uparrow\right\rangle
\right)$,
$\left|\downarrow,\downarrow\right\rangle$ &
$0$ \\
\cline{2-5}

& $1$ &
$\left|S\right\rangle$ &
$\frac{1}{\sqrt{2}}
\left(
\left|\uparrow,\downarrow\right\rangle-
\left|\downarrow,\uparrow\right\rangle
\right)$ &
$2J_\mathrm{H}$ \\
\cline{2-5}

& $1$ &
$\left|D_0\right\rangle$ &
$\sim\left|\uparrow\downarrow,0\right\rangle$ &
$U-J_\mathrm{H}-\sqrt{\Delta^2+J_\mathrm{H}^2}$ \\
\cline{2-5}

& $1$ &
$\left|D_1\right\rangle$ &
$\sim\left|0,\uparrow\downarrow\right\rangle$ &
$U-J_\mathrm{H}+\sqrt{\Delta^2+J_\mathrm{H}^2}$ \\
\hline

\multirow{2}{*}{$3$}
& $2$ &
$\left|2,\sigma\right\rangle$ &
$\left|\uparrow \downarrow,\uparrow\right\rangle,
 \left|\uparrow \downarrow,\downarrow\right\rangle$ &
$0$ \\
\cline{2-5}

& $2$ &
$\left|\sigma,2\right\rangle$ &
$\left|\uparrow, \uparrow \downarrow\right\rangle,
 \left|\downarrow,\uparrow \downarrow\right\rangle$ &
$\Delta$ \\
\hline

$4$ & $1$ &
$\left|2,2\right\rangle$ &
$\left|\uparrow\downarrow,\uparrow\downarrow\right\rangle$ &
$\Delta+J_\mathrm{H}$ \\
\hline

\end{tabular}
\end{table*}

The instantaneous weight of a given local multiplet in the time-evolved state can be quantified by
\begin{equation}
    \rho_P(t)=\frac{1}{L}\sum_\ell\langle \psi(t) |P_{\ell}|\psi(t)\rangle\,.
\end{equation}
The corresponding decomposition is shown in Figs.~\ref{fig::figS5}(a,b). In the ground state ($t=0$), the dominant local configurations are the triplet $|T\rangle$ and the three-electron state $|2,\sigma\rangle$, consistent with the equilibrium filling $n=2.5$. During the pulse, the main change is the transfer of double occupancy from orbital $\gamma=0$ to orbital $\gamma=1$, i.e., the transfer of weight from the $|2,\sigma\rangle$ to the $|\sigma,2\rangle$ multiplet, respectively. In the atomic limit, these three-electron configurations are separated by the crystal-field splitting $\Delta$, providing a natural connection to the two resonant excitation channels identified in the main text: $2\omega_P\simeq\Delta$ for the two-photon process and $\omega_P\simeq\Delta$ for the direct one-photon process.

It is important to note that the two excitation channels show slightly different transient dynamics. For the two-photon excitation shown in Fig.~\ref{fig::figS5}(a), the redistribution between $|2,\sigma\rangle$ and $|\sigma,2\rangle$ is strongly nonmonotonic during the pulse and is accompanied by a small population of additional local configurations, including $|2,2\rangle$. In contrast, direct one-photon excitation in Fig.~\ref{fig::figS5}(b) produces a smoother transfer between the two three-electron configurations.

To characterize not only the instantaneous multiplet content of $|\psi(t)\rangle$, but also the local configurations probed by single-particle excitations $A(q,\omega)$, we further introduce the multiplet-resolved electron and hole part of the density-of-states,
\begin{equation}
\begin{aligned}
\rho^{\mathrm e}_{\gamma,P}(t)&=\frac{1}{L}\sum_{\ell,\sigma}\,\langle\psi(t)|c^\dagger_{\gamma,\ell,\sigma}\, P_\ell\, c_{\gamma,\ell,\sigma}|\psi(t)\rangle\,,\\
\rho^{\mathrm h}_{\gamma,P}(t)&=\frac{1}{L}\sum_{\ell,\sigma}\,\langle\psi(t)| c_{\gamma,\ell,\sigma}\, P_\ell\, c^\dagger_{\gamma,\ell,\sigma}|\psi(t)\rangle\,.
\end{aligned}
\end{equation}
These quantities decompose the DOS according to the local atomic multiplet selected by $P_\ell$. They should be distinguished from $\rho_P(t)$: while $\rho_P(t)$ measures the contribution of a given atomic configuration to the time-evolved wave-function $|\psi(t)\rangle$ itself, $\rho^{\mathrm e/h}_{\gamma,P}(t)$ characterizes the local multiplet content of single-particle excitations.

The resulting decompositions are shown in Figs.~\ref{fig::figS5}(c-f), with panels (c,d) corresponding to orbital $\gamma=0$ and panels (e,f) to orbital $\gamma=1$. Their evolution follows directly from the transfer of the dominant three-electron configuration from $|2,\sigma\rangle$ to $|\sigma,2\rangle$. Initially, removing an electron from $\gamma=0$ gives access to the triplet and singlet sectors, $|2,\sigma\rangle \xrightarrow{c_{\gamma=0}} |T\rangle,|S\rangle$, while removing the electron from $\gamma=1$ leads to $|D_0\rangle$. After the pulse ($t>2t_P$), these channels are interchanged: $|\sigma,2\rangle \xrightarrow{c_{\gamma=1}} |T\rangle,|S\rangle$, while removal from $\gamma=0$ leads mainly to $|D_1\rangle$. Consequently, the triplet and singlet densities decrease in orbital $\gamma=0$ and increase in orbital $\gamma=1$, while the dominant doublon contribution changes from $|D_0\rangle$ to $|D_1\rangle$. Since $|2,\sigma\rangle$ and $|\sigma,2\rangle$ are separated by the crystal-field energy $\Delta$, the transitions from these configurations to the same two-electron multiplet differ in energy by $\Delta$.

\section{One- and two-photon dynamics in the $L=2$ system}
\label{sec:L2_driving}
We analyze the two excitation channels using the eigenstates of a field-free $L=2$ system and a constant amplitude monochromatic drive, $A(t)=a_0\cos(\omega_Pt)$. The exact diagonalization in the $N_\uparrow=3$, $N_\downarrow=2$ ($n=2.5$ as in the main text) sector gives 24 states. The spin-independent Peierls drive preserves $S=3/2$, restricting dynamics from the ground state to the four states listed in Table~\ref{tab:L2_states}. We denote the ground state by $|g\rangle$, the ground-state-like even-parity state by $|i\rangle$, and the two orbital-transfer states by $|e_1\rangle$ and $|e_2\rangle$ (the subscripts indicate the photon order of their excitation channels). Although $|i\rangle$ is one-photon active, its orbital occupations are nearly identical to those of $|g\rangle$. In contrast, $|e_1\rangle$ and $|e_2\rangle$ have orbital occupations associated with charge inversion transfer.

\begin{table}[ht!]
\caption{Relevant eigenstates for the $L=2$ system. Energies are measured relative to the ground state. The subscripts on $e_1$ and $e_2$ indicate the photon order of the orbital-transfer channel.}
\label{tab:L2_states}
\centering
\resizebox{\columnwidth}{!}{%
\begin{tabular}{ccccccc}
\hline\hline
State & $E_s-E_g$ & Parity & $\langle n_0\rangle$ & $\langle n_1\rangle$ & $|J_{sg}|$ & $|K_{sg}|$ \\
\hline
$|g\rangle$   & $0$        & $-1$ & $1.499856$ & $1.000144$ & $0$        & $0.503273$ \\
$|i\rangle$   & $0.999734$ & $+1$ & $1.499808$ & $1.000192$ & $0.499534$ & $0$        \\
$|e_2\rangle$ & $5.903389$ & $-1$ & $1.000144$ & $1.499856$ & $0$        & $0.093167$ \\
$|e_1\rangle$ & $6.103654$ & $+1$ & $1.000192$ & $1.499808$ & $0.111490$ & $0$        \\
\hline\hline
\end{tabular}%
}
\end{table}

With $J=\left.\partial_AH\right|_{A=0}$ and $K=\left.\partial_A^2H\right|_{A=0}$, expansion of the Peierls Hamiltonian gives
\begin{equation}
H(t)=H_0+a_0J\cos(\omega_Pt)
+\frac{a_0^2}{4}K[1+\cos(2\omega_Pt)]
+O(a_0^3).
\label{eq:S_Peierls_expansion}
\end{equation}
The operator $J$ is odd under inversion, and $K$ is even. Thus, $J$ connects the odd-parity ground state $|g\rangle$ to the even-parity state $|e_1\rangle$, while the odd-parity state $|e_2\rangle$ can be reached through $K$ or through two successive $J$ couplings via the even-parity states $|i\rangle$ and $|e_1\rangle$. In the following, we use the notation $K_{\alpha\beta}=\langle \alpha |K|\beta\rangle$ and $J_{\alpha\beta}=\langle \alpha |J|\beta\rangle$.

For a nearly isolated $n$-photon resonance, with $n=1,2$, we simply denote the corresponding target state $|e_n\rangle$ by $|e\rangle$ and project the dynamics onto $|g\rangle$ and $|e\rangle$. In this basis the effective rotating-frame Hamiltonian is given by
\begin{equation}
H_{\mathrm{eff}}^{(n)}=
\begin{pmatrix}
0 & g_n^*\\
g_n & \delta_n
\end{pmatrix},
\label{eq:S_eff_hamiltonian}
\end{equation}
where $\delta_n=E_e-E_g-n\omega_P+\Delta E_e-\Delta E_g$, and $\Delta E_\alpha$ are the quasienergy shifts induced by the drive of the ground and target states. Consequently, setting $\omega_P=(E_e-E_g)/n$ cancels the field-free detuning, but generally does not yield $\delta_n=0$.

For the one-photon channel, the first harmonic in Eq.~\eqref{eq:S_Peierls_expansion} gives
\begin{equation}
g_1=\frac{a_0}{2}J_{e_1g},\qquad \Omega^{(1)}=2|g_1|\simeq0.1115a_0.
\label{eq:S_one_coupling}
\end{equation}
In the two-photon channel, the $2\omega_P$ harmonic of $K$ provides a direct coupling, while two successive $J$ couplings act through $|i\rangle$ and $|e_1\rangle$:
\begin{equation}
\begin{aligned}
g_2={}&\frac{a_0^2}{8}K_{e_2g} +\frac{a_0^2}{8}\sum_{m=i,e_1}J_{e_2m}J_{mg}\\
&\times\left(\frac{1}{E_g+\omega_P-E_m}+\frac{1}{E_{e_2}-\omega_P-E_m}\right).
\end{aligned}
\label{eq:S_two_coupling}
\end{equation}
The direct and virtual contributions add coherently. Evaluated near the field-free two-photon resonance, they give $\Omega^{(2)}=2|g_2|\simeq0.0330a_0^2$.

For either state $|\alpha\rangle=|g\rangle$ or $|e\rangle$, the quasienergy shift $\Delta E_\alpha$ up to the second order in $a_0$ is given by
\begin{equation}
\Delta E_\alpha=
\frac{a_0^2}{4}K_{\alpha\alpha}
+\frac{a_0^2}{4}\sum_m\sum_{q=\pm1}
\frac{|J_{m\alpha}|^2}{E_\alpha-E_m+q\omega_P}.
\label{eq:S_level_shift}
\end{equation}
Here, $m$ runs over the four states in Table~\ref{tab:L2_states}, and $q=\pm1$ labels the two Fourier components of the linear drive. Note that some terms are already included in the effective two-level Hamiltonian [Eq.~\eqref{eq:S_eff_hamiltonian}] and should be excluded from the sum: at the one-photon resonance, $(m,q)=(e_1,+1)$ for $s=g$ and $(m,q)=(g,-1)$ for $s=e_1$. No term is excluded at the two-photon resonance. The first term comes from the time-independent part of the quadratic Peierls coupling, whereas the sum is the second-order shift generated by the oscillating $J$ coupling. The detunings $\delta_n$ becomes
\begin{equation}
\begin{aligned}
\delta_1&\simeq E_{e_{1}}-E_g-\omega_P-0.1548a_0^2,\\
\delta_2&\simeq E_{e_{2}}-E_g-2\omega_P-0.1151a_0^2.
\end{aligned}
\label{eq:S_detunings}
\end{equation}
The coefficients of $a_0^2$ are evaluated at the corresponding field-free frequencies. Their variation under a frequency shift of order $a_0^2$ contributes only in higher order.

Starting from $|g\rangle$, the effective two-level model predicts
\begin{equation}
\begin{aligned}
p_e(t)&=|\langle e|\psi(t)\rangle|^2\\
&\simeq
\frac{(\Omega^{(n)})^2}{(\Omega^{(n)})^2+\delta_n^2}
\sin^2\!\left[
\frac{t}{2}\sqrt{(\Omega^{(n)})^2+\delta_n^2}
\right].
\end{aligned}
\label{eq:S_population}
\end{equation}
Setting $\delta_2\simeq0$ in Eq.~\eqref{eq:S_detunings} gives,
to second order in $a_0$,
\begin{equation}
\omega_{P,\mathrm{res}}^{(2)}
\simeq\frac{E_{e_2}-E_g}{2}-0.05754a_0^2.
\label{eq:S_shifted_resonance}
\end{equation}

At $a_0=0.1$, the drive-induced shift of the two-photon pump resonance is $\omega_{P,\mathrm{res}}^{(2)}-\omega_P\simeq-5.75\times10^{-4}$. Despite this small change in the pump frequency, tuning to the shifted resonance restores nearly complete population exchange, as shown by the dashed curves in Fig.~\ref{fig::fig3}(b) of the main text. Its much longer period follows from $\Omega^{(2)}\ll\Omega^{(1)}$ at the same weak-drive amplitude. 

\section{Long-time behavior}
\begin{figure}[b!]
\centering
\includegraphics[width=0.8\columnwidth]{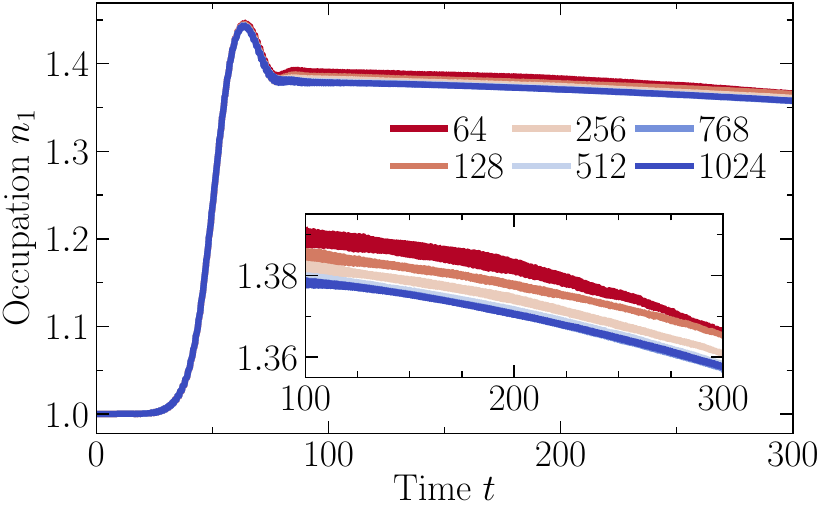}
\caption{Long-time evolution of $n_1$ for MPS bond dimensions $\chi=64$, $128$, $256$, $512$, $768$, and $1024$ with $L=16$ and $\Delta=5.5$. The system is driven by a non-chirped pulse with $a_0=2.2$ and $\omega_P=2.8$, addressing the two-photon excitation channel. The inset enlarges the post-pulse regime, $100\leq t\leq300$, showing only weak bond-dimension dependence and slow relaxation.}
\label{fig::figS6}
\end{figure}

To verify convergence with respect to the MPS bond dimension $\chi$, we compare the time evolution of $n_1$ for $\chi=64$, $128$, $256$, $512$, $768$ and $1024$. As shown in Fig.~\ref{fig::figS6}, increasing $\chi$ leads only to small quantitative changes in the post-pulse evolution, while the results for the largest bond dimensions are nearly converged. Post-pulse occupation remains close to its plateau value throughout the accessible time window, up to $t=300$, supporting the long-lived character of the inverted OSMP. Based on this convergence analysis, we use $\chi=512$ in the main-text calculations.

\end{document}